\documentclass[conference]{IEEEtran}

\usepackage[font=normal]{caption}
\usepackage{subcaption}
\usepackage{enumitem}
\usepackage{amssymb}
\usepackage{tabularx} 
\IEEEoverridecommandlockouts
\usepackage{tabularray}
\usepackage{url}
\usepackage{tcolorbox}
\usepackage{listings}
\usepackage{xcolor}
\usepackage[export]{adjustbox}
\usepackage{booktabs}
\usepackage{graphicx}
\usepackage{amsmath}
\usepackage{balance}
\usepackage{fancyhdr}
\usepackage{multirow}
\usepackage{hyperref}
\hypersetup{
    colorlinks=true,
    linkcolor=blue,
    filecolor=blue,      
    urlcolor=blue,
    citecolor=blue,
    pdfpagemode=FullScreen,
    }
\usepackage{listings}

\usepackage{xcolor}
\usepackage{soul}

\definecolor{codegreen}{rgb}{0,0.6,0}
\definecolor{codegray}{rgb}{0.5,0.5,0.5}
\definecolor{codepurple}{rgb}{0.58,0,0.82}
\definecolor{backcolour}{rgb}{0.95,0.95,0.92}

\lstdefinestyle{mystyle}{
  label=code:sample,
  floatplacement=tbp,
  backgroundcolor=\color{backcolour}, commentstyle=\color{codegreen},
  keywordstyle=\color{magenta},
  numberstyle=\tiny\color{codegray},
  stringstyle=\color{codepurple},
  basicstyle=\ttfamily\footnotesize,
  breakatwhitespace=false,         
  breaklines=true,                 
  captionpos=t,                    
  keepspaces=true,                 
  numbers=left,                    
  numbersep=2pt,                  
  showspaces=false,                
  showstringspaces=false,
  showtabs=false,                  
  tabsize=1
}

\usepackage{graphicx}
\usepackage{array}
\usepackage{multirow}
\usepackage{colortbl}
\usepackage{rotating}
\AtBeginDocument{%
  \providecommand\BibTeX{{%
    \normalfont B\kern-0.5em{\scshape i\kern-0.25em b}\kern-0.8em\TeX}}}

\begin{document}

\title{Toward Realistic Evaluations of Just-In-Time Vulnerability Prediction}

\author{
    \IEEEauthorblockN{Duong Nguyen\IEEEauthorrefmark{1}, Thanh Le-Cong\IEEEauthorrefmark{2}, Triet Huynh Minh Le\IEEEauthorrefmark{3}, M. Ali Babar\IEEEauthorrefmark{3}, Quyet-Thang Huynh\IEEEauthorrefmark{1}}
    \IEEEauthorblockA{\IEEEauthorrefmark{1}\textit{School of Communication and Information Technology, Hanoi University of Science and Technology, Hanoi, Vietnam}
    \\ duong.nd215336@sis.hust.edu.vn, thang.huynhquyet@hust.edu.vn}
    \IEEEauthorblockA{\IEEEauthorrefmark{2}\textit{School of Computing and Information Systems, The University of Melbourne, Melbourne, Australia}
    \\ congthanh.le@student.unimelb.edu.au}
    \IEEEauthorblockA{\IEEEauthorrefmark{3}\textit{School of Computer and Mathematical Sciences, The University of Adelaide, Adelaide, Australia}
    \\ \{triet.h.le, ali.babar\}@adelaide.edu.au}
}

\maketitle
\thispagestyle{fancy}

\cfoot{\thepage} 
\renewcommand{\headrulewidth}{0pt} 
\renewcommand{\footrulewidth}{0pt}
\pagestyle{fancy}
\cfoot{\thepage} 

\begin{abstract}
Modern software systems are increasingly complex, presenting significant challenges in quality assurance. Just-in-time vulnerability prediction (JIT-VP) is a proactive approach to identifying vulnerable commits and providing early warnings about potential security risks. However, we observe that current JIT-VP evaluations rely on an idealized setting, where the evaluation datasets are artificially balanced, consisting exclusively of vulnerability-introducing and vulnerability-fixing commits. 

To address this limitation, this study assesses the effectiveness of JIT-VP techniques under a more realistic setting that includes both vulnerability-related and vulnerability-neutral commits. To enable a reliable evaluation, we introduce a large-scale public dataset comprising over one million commits from FFmpeg and the Linux kernel. Our empirical analysis of eight state-of-the-art JIT-VP techniques reveals a significant decline in predictive performance when applied to real-world conditions; for example, the average PR-AUC on Linux drops 98\% from 0.805 to 0.016. This discrepancy is mainly attributed to the severe class imbalance in real-world datasets, where vulnerability-introducing commits constitute only a small fraction of all commits.

To mitigate this issue, we explore the effectiveness of widely adopted techniques for handling dataset imbalance, including customized loss functions, oversampling, and undersampling. Surprisingly, our experimental results indicate that these techniques are ineffective in addressing the imbalance problem in JIT-VP. These findings underscore the importance of realistic evaluations of JIT-VP and the need for domain-specific techniques to address data imbalance in such scenarios.
\end{abstract}

\section{Introduction}
\label{sec:intro}
Software vulnerabilities pose significant risks to the reliability, security, and functionality of modern software systems, often leading to severe consequences. 
For example, the 2024 CrowdStrike outage illustrates how a seemingly minor misalignment between desired behaviors and their actual implementations caused a cascade of system failures, affecting millions of devices and disrupting essential services worldwide~\cite{crowdstrike,techtarget}. Such incidents highlight not only the technical risks embedded within modern software systems, but also the significant financial and operational burdens incurred when vulnerabilities are detected post-deployment.

In response to these challenges, Just-In-Time Vulnerability Prediction (JIT-VP)~\cite{perl2015vccfinder} has emerged as a promising approach to improve software quality assurance. JIT-VP techniques aim to identify potential security threats at the point of code commit, thereby providing immediate and actionable feedback to developers during the early stages of the software development life cycle. By learning from historical vulnerability-introducing commits, these methods can effectively flag potentially vulnerable code modifications, reducing the cost and effort associated with subsequent remediation. Consequently, the integration of JIT-VP into the development life cycle not only improves the efficiency of security inspections but also strengthens the overall resilience of software systems against emerging threats. Recent studies~\cite{nguyen2024code, perl2015vccfinder, riom2021revisiting} have demonstrated promising performance in Just-In-Time Vulnerability Prediction (JIT-VP). For example, CodeJIT~\cite{nguyen2024code} can successfully detect 80\% of vulnerability-introducing commits of known vulnerabilities with a precision of 90\%. 

Despite these encouraging results, we identify potential deficiencies in the construction of the benchmark datasets utilized by these studies. Specifically, existing studies~\cite{nguyen2024code}\cite{perl2015vccfinder}\cite{riom2021revisiting} have purely restricted their datasets to vulnerability-related commits, including \textit{Vulnerability-Introducing Commits} (VICs) as vulnerable commits and \textit{Vulnerability-Fixing Commits} (VFCs) as safe commits. Unfortunately, in practical applications, the input of JIT-VP models may include not only vulnerability-related commits but also other commits that are irrelevant to introducing and/or fixing vulnerabilities, which we hereafter refer to as \textit{Vulnerability-Neutral Commits} (VNCs). When deploying JIT-VP tools in practice, filtering out VNCs from the commit set is infeasible. Therefore, assuming their absence in evaluation settings is unrealistic.
Consequently, training and evaluating JIT-VP techniques solely on the restricted categories of VICs and VFCs lead to an incomplete assessment of their performance in practical scenarios. This limitation raises concerns regarding the real-world effectiveness of JIT-VP techniques, as their ability to handle a more representative and diverse distribution of commits remains largely unexplored.

To address these concerns, we conduct a large-scale empirical study to evaluate and compare the effectiveness of JIT-VP techniques in two settings: (1) the \textit{idealized} setting,  and (2) the \textit{realistic} setting. Specifically, in the idealized setting, the dataset is restricted to only VICs and VFCs, as widely used in previous studies. Meanwhile, the realistic setting incorporates VFCs, VICs, and all remaining VNCs, which more accurately reflects real-world scenarios. We first construct datasets of 1 million commits for these settings from two well-known open-source projects: FFmpeg~\cite{ffmpeg} and the Linux kernel~\cite{linux}. Then, leveraging these datasets, we systematically evaluate eight state-of-the-art approaches for JIT-VP in these two settings.

Our experiments shed light on the JIT-VP performance in both idealized and realistic settings.
In the idealized setting, we demonstrate that existing techniques yield impressive performance, with an average PR-AUC of over 0.8. Notably, the best-performing existing baseline, JIT-Fine~\cite{ni2022best}, achieves a PR-AUC of 0.96 on FFmpeg. However, when these techniques are applied in the realistic setting, their performance declines remarkably, with the average PR-AUC decreasing by over 90\% from more than 0.8 to 0.091 and 0.016 on FFmpeg and Linux, respectively. Specifically, the performance of JIT-Fine, the best baseline in the idealized setting, decreases substantially to a PR-AUC of 0.111 in FFmpeg and drops to only 0.005 on Linux. These findings strongly indicate that the current JIT-VP evaluation settings are overly idealized and may lead to misleading conclusions about their practical effectiveness.

\begin{figure}[t]
    \centering
    \includegraphics[width=\linewidth, trim=10 10 10 10,clip]{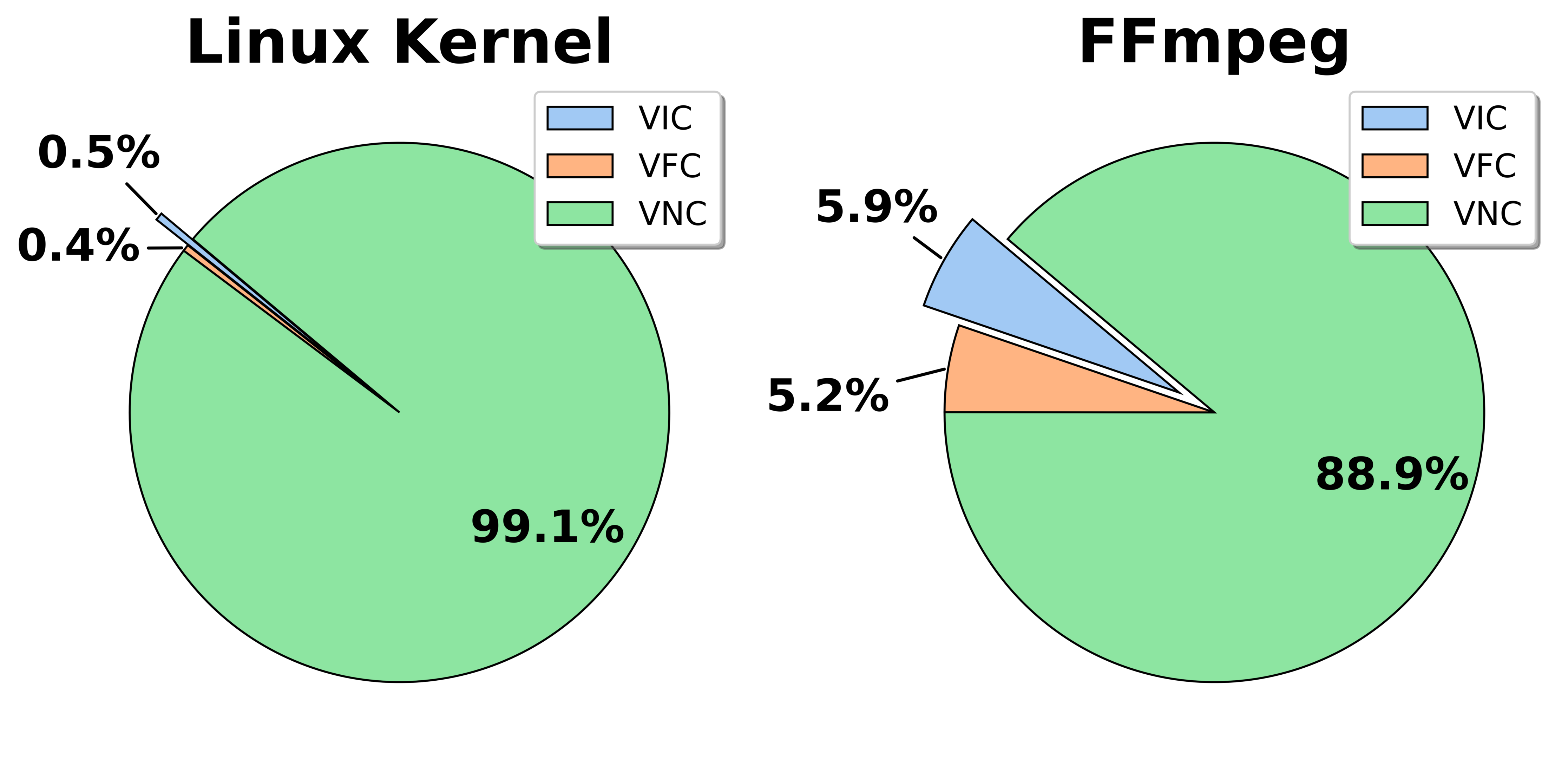}
    \caption{Distributions of different types of commits in the Linux and FFmpeg projects. VIC, VFC, VNC denotes vulnerability-introducting, vulnerability-fixing and vulnerability-neural commits, respectively.}
    \label{fig:distribution}
\end{figure}

We found that these performance declines are caused by the extreme imbalance in the realistic dataset. Specifically, as illustrated in Figure~\ref{fig:distribution}, VNCs account for more than 90\% of all commits on both FFmpeg and Linux. Therefore, the absence of VNCs in the idealized setting leads to an artificially balanced ratio between safe and dangerous commits. For instance, the vulnerable-to-safe commit ratios in this setting are 1:0.89 for FFmpeg and 1:0.88 for Linux. However, in real-world scenarios, these ratios change drastically to approximately 1:17 and 1:217, respectively. This extreme imbalance poses significant challenges for machine learning models to achieve accurate predictions for the JIT-VP task.

To address this challenge, we evaluate the effectiveness of five widely used approaches to mitigate imbalance in machine learning models: customized loss functions (e.g., FLoss~\cite{lin2017focal}), random oversampling, random undersampling, SMOTE~\cite{fernandez2018smote}, and OSS~\cite{kubat1997addressing}. Unfortunately, our experimental results demonstrate that these methods not only fail to enhance the performance of JIT-VP techniques but, in some cases, even degrade it.
These findings highlight the challenges of addressing the imbalance in the JIT-VP dataset, emphasizing the necessity of advanced and domain-specific methods to tackle this problem.

In summary, the \textit{contributions} of this paper are as follows:

\begin{itemize}
    \item \textbf{Large-scale Dataset}: We release a publicly available dataset of over one million commits, including their code changes and human-engineered features, collected from two well-known open source projects: FFmpeg and the Linux kernel. To the best of our knowledge, this is the largest dataset currently available for JIT-VP. The data set is available in our replication package~\cite{replication}.
    
    \item \textbf{Comprehensive Evaluations}: We extensively evaluate state-of-the-art JIT-VP approaches under the idealized setting and realistic settings, thereby highlighting the gaps between controlled and real-world performance.
    
    \item \textbf{Investigation of Imbalance Mitigation}: We analyze the challenges introduced by the imbalance of the dataset in the realistic setting and assess the effectiveness of commonly used data imbalance mitigation techniques. Our results show limited effectiveness of these existing techniques, emphasizing the need for advanced domain-specific solutions for JIT-VP.
\end{itemize}

\textbf{Data Availability.} To support the open science initiative, we publish our replication package at \href{https://figshare.com/s/bedbc45f494aed760e06}{https://figshare.com/s/bedbc45f494aed760e06}. Moreover, we also release VulGuard, a tool developed as part of this study at \href{https://github.com/AI4Code-HUST/VulGuard/}{https://github.com/AI4Code-HUST/VulGuard/}.

\section{Background and Motivation}
\label{sec:background}
\subsection{Just-In-Time Vulnerability Prediction}
Due to the critical impact of software vulnerabilities, the automatic identification of potential vulnerabilities has gained significant attention with various research directions including dynamic testing~\cite{li2018vuldeepecker}, static program analysis~\cite{4362893, ferschke2012flawfinder, zhou2024comparison} and machine learning~\cite{li2018vuldeepecker, chakraborty2021deep, li2021sysevr,le2024software}. However, these approaches mainly perform their analyses at the code file/function level, often detecting vulnerabilities only after deployment.

Software systems witness frequent changes in practice, necessitating early feedback on commits or pull requests. To address this demand, a common approach is code review, in which human developers manually review code commits to identify potential security vulnerabilities. However, this process is both time-consuming and labor-intensive. Recently, Just-In-Time Vulnerability Prediction (JIT-VP) has emerged as a promising solution to mitigate these challenges by prioritizing high-risk commits, i.e., those that have changes likely to introduce vulnerabilities. Specifically, JIT-VP methods aim to identify such commits by analyzing associated metadata, including code changes, commit messages, and authors' information. More formally, the JIT-VP task is defined as follows:
\begin{tcolorbox}
Given a commit $c$ and its associated information $\mathcal{I}(c)$, define a mapping function $f$ so that:
\begin{equation*}
  f(\mathcal{I}(c))=
  \begin{cases}
     1 &\text{if $c$ is vulnerable}\\
     0 &\text{if $c$ is safe}
  \end{cases}
\end{equation*}
\end{tcolorbox}

JIT-VP approaches commonly utilize machine learning models to learn this mapping function $f$ from historical data and then use the trained model to predict future commits. Perl et al.~\cite{perl2015vccfinder} proposed VCCFinder, a Support Vector Machine model, for JIT-VP using hand-crafted features and messages. As the VCCFinder replication package is not available, Riom et al. replicated this technique and released their implementation as a reproducible baseline, which has been adopted in many subsequent works~\cite{nguyen2024code, chen2024improvingdatacurationsoftware}. Yang et al. proposed VulDigger~\cite{yang2017vuldigger}, a cost-aware JIT-VP model combining a classification component to detect vulnerable commits and a regression model to pinpoint suspicious changes with minimal inspection effort. Other approaches, such as CodeJIT~\cite{nguyen2024code} and VDA~\cite{li2023commit}, leverage graph-based representations to further improve the performance of JIT-VP. 
We use these existing techniques in our experiments; more details can be found in Section~\ref{sec:models}.

Empirical work conducted by Lomio et al.~\cite{lomio2022just} is the most closely related to our research. The study conducts its analysis in a cross-project setting, in which commits from multiple software projects are utilized to train models to distinguish vulnerable commits from benign ones. However, the findings are limited by both the small number of known vulnerabilities and computational constraints, which hinder the generalization of the results to real-world vulnerability prediction scenarios. Specifically, the dataset includes only 90 vulnerable commits, with the negative (non-vulnerable) samples down to maintain a fixed 1:100 positive-to-negative ratio. As a result, projects with few reported vulnerabilities contribute disproportionately fewer total commits to the training set, leading to under-representation of "safe" commits in some cases. This imbalance may impair the model's ability to generalize across diverse project contexts. In contrast, our study not only covers more vulnerabilities, but also encompasses both an idealized scenario and a more realistic setting characterized by imbalanced datasets, as detailed in Section~\ref{sec:data_col_limits}.
Furthermore, we extend beyond the general purpose machine learning models employed in~\cite{lomio2022just} by incorporating domain-specific models tailored for both JIT defect prediction and vulnerability prediction, allowing for a more comprehensive evaluation.

\subsection{Existing Data Collection for Just-In-Time Vulnerability Prediction and Its Limitations}
\label{sec:data_col_limits}
A large volume of high-quality data is fundamental to training high-performance learning models. This subsection introduces common data collection approaches for JIT-VP. 

\vspace{2mm}

\textbf{Vulnerability Identification in Commits}. The first step of data collection is to identify commits that introduce vulnerabilities into the software, known as \textit{\textbf{Vulnerability-Introducing Commits (VICs)}}. However, collecting VICs remains a significant challenge~\cite{perl2015vccfinder}. A common approach is to begin with \textit{\textbf{Vulnerability-Fixing Commits (VFCs)}}, the commits that patch or fix vulnerabilities, since they are often well-documented and exhibit distinctive features in their messages and code changes. Numerous studies have explored automated approaches~\cite{nguyen2022vulcurator, nguyen2023midas} to identify silent VFCs from open source projects. Although these methods provide the advantage of uncovering a broader range of commits, including undocumented ones, they also introduce a high risk of false positives. Consequently, extracting VFCs from CVE/NVD entries~\cite{nguyen2025mapping, sabetta2024known, li2024patchfinder} or relying on human judgment~\cite{zhou2019devign} remains a preferred strategy to ensure accuracy and reliability. After identifying VFCs, VICs are typically traced using SZZ algorithms~\cite{sliwerski2005changes, bao2022v}, which link fixes to their root causes by tracking code modifications. SZZ is widely used in vulnerability analysis and defect prediction as it can effectively identify commits that introduce vulnerabilities, but it still has limitations, such as attributing the blame primarily to VFCs rather than newly added lines~\cite{rodriguez2018reproducibility, bao2022v}. Human annotators are often considered another option to ensure reliability, despite being time-consuming and expensive~\cite{zhou2019devign,liu2020large}. Recent studies have proposed an alternative identification technique that relies on developers' commit messages or patch notes to pinpoint the source of the weakness~\cite{rosa2021evaluating,lyu2024evaluating}.

\vspace{2mm}

\textbf{Limitations of Current Data Labeling}. Accurate data labeling plays a crucial role in mitigating biases, reducing noise, and enhancing model generalizability, ultimately leading to more reliable and effective predictions~\cite{croft2023data}. In vulnerability-related commit classification, Vulnerability-Introducing Commits (VICs) are labeled as positive instances, while ``safe'' commits are typically classified as Vulnerability-Fixing Commits (VFCs).
However, the current datasets for the task remain relatively small and balanced.
For example, in CodeJIT~\cite{nguyen2024code}, the curated dataset consisted of 8,975 VICs and 11,299 VFCs.
In addition, VCCFinder~\cite{perl2015vccfinder}, despite having collected a large dataset of 170,860 commits from 66 repositories, trained and evaluated its model on a significantly smaller subset of only 1,258 commits, restricted to VICs and VFCs.

As demonstrated in Figure~\ref{fig:distribution}, vulnerability-related commits, including VICs and VFCs, account for only approximately 10\% and 1\% of the total commits in the FFmpeg and Linux kernel projects, respectively. Consequently, in practical scenarios, a given commit is more likely to belong to neither the VIC nor VFC categories, thereby increasing the complexity of classification compared to the current experimental setting. Therefore, to improve the practicality of evaluating JIT-VP, it is crucial to incorporate the remaining commits that are not known to introduce or fix vulnerabilities, referred to as \textit{\textbf{Vulnerability-Neutral Commits (VNC)}}, into the evaluations. 

To address this challenge, we propose a new evaluation framework for JIT-VP, termed the \textbf{\textit{realistic setting}}, which better reflects real-world scenarios. In the realistic setting, the evaluation dataset comprises VICs, VFCs, and VNCs. Within this dataset, VICs are labeled as vulnerable commits, while both VFCs and VNCs are categorized as safe commits.
In contrast, we define the commonly used evaluation approach, where only VICs and VFCs are considered, as the \textbf{\textit{idealized setting}}. In the idealized setting, the evaluation dataset consists exclusively of VICs and VFCs, where VICs are labeled as vulnerable commits and VFCs as safe commits.
To the best of our knowledge, we are the first to investigate the gaps between these two settings, aiming to enhance the applicability and reliability of the JIT-VP evaluation and ensure that it aligns more closely with the conditions encountered in real-world software development.

\section{Study Design}
\label{sec:method}

\subsection{Research Questions}

Our empirical study aims to answer the following Research Questions (RQs):

\textbf{RQ$_{1}$}. \textit{How effective are the state-of-the-art techniques in just-in-time vulnerability prediction in the idealized setting used by previous studies?} In this RQ, we evaluate the latest JIT-VP techniques in the idealized setting, an experimental setting widely adopted in existing studies~\cite{nguyen2024code, perl2015vccfinder}. Our investigation has two primary objectives. First, we validate previous findings using our dataset, which incorporates the latest data and a broader range of JIT-VP techniques. Second, the results of this RQ are expected to serve as a baseline for comparison with those in the realistic setting in the subsequent RQ.

\textbf{RQ$_{2}$}. \textit{How effective are state-of-the-art techniques in just-in-time vulnerability prediction in the realistic setting?} In this RQ, we evaluate existing JIT-VP approaches using our proposed realistic setting, which aims to accurately reflect real-world scenarios where all commits within a project must be classified. Our objectives are two-fold: (1) to assess the true effectiveness of JIT-VP techniques in practice and (2) to analyze the gap in performance between the idealized setting and the more complex, real-world realistic setting.

\textbf{RQ$_{3}$}. \textit{How effective are commonly used techniques in mitigating data imbalance observed in the realistic setting?} As we will show in Section~\ref{sec:rq2}, the performance of JIT-VP techniques declines significantly in the realistic setting due to the data imbalance inherent in real-world datasets. In this RQ, we investigate the effectiveness of widely used methods for addressing this issue, including weighted loss functions, oversampling, and undersampling. We aim to shed light on the extent to which current techniques can mitigate this fundamental challenge, thereby informing the development of better JIT-VP techniques to be deployed in practice.

\subsection{Data Collection}~\label{sec:data_collection}
\subsubsection{Project Selection}
To construct a high-quality dataset, it is essential to first identify and select reliable and well-maintained projects. In this study, we chose FFmpeg and the Linux kernel as our data sources for two reasons.

\textbf{First, they are large-scale projects.} Both FFmpeg and Linux are well-known open-source projects with extensive histories. As a result, they attract significant contributions from developers worldwide, leading to a substantial volume of commits. Specifically, as of September 24, 2024, FFmpeg and Linux have accumulated approximately 117,000 and 1,300,000 commits, respectively. The substantial number of commits in these projects serves as a crucial enabler for developing a high-quality dataset. 
    
\textbf{Second, both FFmpeg and Linux are well maintained and have vulnerability reporting practices}, including security guidelines and comprehensive vulnerability and patch reports. For FFmpeg, there is an official website~\cite{FFmpegWebsite} containing all available documentation, along with a Wiki page~\cite{FFmpegWiki} that archives up-to-date bug reports and patches. The Linux kernel also has a comprehensive documentation website~\cite{LinuxKernelWebsite} and a mailing list with up-to-date CVEs~\cite{LinuxCVE}. These vulnerability reporting practices ensure that datasets extracted from these projects accurately reflect the realistic distribution of vulnerable and non-vulnerable commits.

\subsubsection{Commit Curation} For each project, we scraped all commits up to September 24, 2024. Since FFmpeg and the Linux kernel are large-scale open-source software projects with multiple releases, we only used the default branch \textit{master} that has the latest production code. Following the previous works~\cite{kim2006automatic, mcintosh2018fix}, we discarded all merged commits, whitespace commits, and comment commits. We also removed all commits that did not contain C/C++ code files (.c/.cpp extension) or header files (.h extension) changes. These filtering steps ensure that all commits in the dataset introduce main modifications in our target projects since both FFmpeg and Linux are mainly implemented in C/C++. Note that these pre-processing steps are based on known information available at the time of prediction, without prior knowledge of vulnerabilities. Thus, these filtering processes do not compromise the realistic distribution of commits, unlike removing VNCs, which would require knowledge of the vulnerabilities to be filtered.

Subsequently, each of the curated commits was annotated as either VFC, VIC, or VNC by adopting the commonly used data curation pipeline mentioned in Section \ref{sec:data_col_limits}. We started with VFCs and traced back code modifications to identify VICs. The remaining commits were then flagged as VNCs because there is no perfect security testing or labeling in practice, and we could only assume that those commits do not introduce or fix vulnerabilities, as per current practice (e.g.,~\cite{jimenez2019importance,croft2022noisy}). The collection of VFCs and VICs is described below.

\vspace{2mm}

\textbf{Vulnerability Fixing Commits.}
Specifically, we first collected VFCs from two well-known datasets proposed by Liu et al.~\cite{liu2020large} and Zhou et al.~\cite{zhou2019devign}
VFC datasets from these studies have been manually annotated, ensuring their reliability. Then, all VFCs of FFmpeg and Linux in these datasets were retrieved and subjected to the same filtering procedure as the aforementioned commit scraping step to remove all non-code changes and unrelated commits. For example, we discovered a commit in the FFmpeg repository\footnote{https://github.com/ffmpeg/FFmpeg/commit/8ef2c79} that was flagged as a VFC in the Devign dataset~\cite{zhou2019devign} although it only contained a change of a text file in the \texttt{doc} subdirectory. 

\begin{figure}[t]
    
    \centering
    \includegraphics[width=0.6\columnwidth]{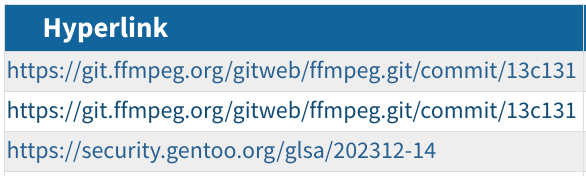}
    \caption{CVE-2022-3965~\cite{CVE-2022-3965} references. The first url links to commit 13c131 in the git.ffmpeg.org repository, which can be traced to its mirror on GitHub.}
    \label{fig:cve-ref}
\end{figure}

Next, we further extended the VFCs collections by examining CVE/NVD~\cite{NVD} entries and official vulnerability reports. Since 90\% of the CVE entries did not refer to GitHub commit links~\cite{liu2020large}, we actively searched for commit-like strings instead. For example, in CVE-2022-3965~\cite{CVE-2022-3965} in Figure \ref{fig:cve-ref}, there are three references in this entry. Unfortunately, none of these links leads to a GitHub page. However, the first URL points to \texttt{git.ffmpeg.org}, the actual repository of the FFmpeg project, so that we can identify the mirror commit in the GitHub repository using the extracted ID from this link. Inspired by this example, we fetched all CVE entries that mention the FFmpeg or Linux kernel projects in their descriptions. Then, we checked their references to remove unrelated domain names and kept commit-like strings with a length of nine or more characters. After that, these commits would be traced to identify their GitHub commit counterparts. Our study only focuses on \textit{master} branch commits. However, we observed that many CVE entries refer to commits of release branches instead of the main production branch. Hence, all commits that could not be traced to their upstream equivalents were removed.

\begin{figure}[t]
    
    \centering
    \includegraphics[width=\columnwidth]{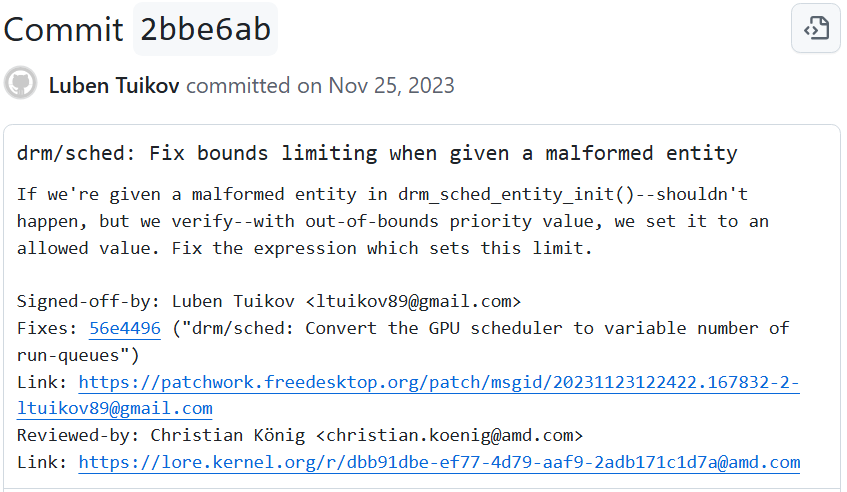}

    \caption{CVE-2023-52461~\cite{CVE-2023-52461} upstream patch commit. Its message state to fix commit 56e4496. The introducing commit can be traced in the master branch.}
    \label{fig:linuxfix}
\end{figure}
 
\begin{table}[t]
\centering
\caption{Summary of Commit Data Statistics for FFmpeg and the Linux kernel. The table presents the number of vulnerability-introducing commits (\#VIC), vulnerability-fixing commits (\#VFC), vulnerability-neural commits (\#VNC), and the total number of commits in the training, validation, and testing sets.}
\label{data:stat}
\begin{tabular}{@{}llccccc@{}}
\hline
\textbf{Project} & \textbf{Partition} & \textbf{\#VIC} & \textbf{\#VFC} & \textbf{\#VNC} & \textbf{\#Total} \\ \hline
\multirow{3}{*}{\textbf{FFmpeg}}       
    & \text{Training}  & 3,826 & 2,519 & 39,823  & 43,650  \\
    & \text{Validation} & 255   & 330   & 3,242   & 3,827   \\
    & \text{Testing}  & 1,020 & 1,903 & 37,778  & 40,701  \\ \hline
\multirow{3}{*}{\textbf{Linux Kernel}} 
    & \text{Training}  & 3,461 & 1,735 & 796,965 & 800,426 \\
    & \text{Validation} & 231   & 616   & 35,086  & 35,317  \\
    & \text{Testing}  & 922   & 1,691 & 157,039  & 157,961  \\ \hline
\textbf{All Commits} & & 9,715  & 8,996  & 1,069,933 & 1,081,882 \\ \hline
\end{tabular}
\end{table}

\vspace{2mm}

\textbf{Vulnerability-Introducing Commits}. Given the collected VFCs, we used the V-SZZ algorithm \cite{bao2022v}, which is an improvement over the traditional SZZ algorithm~\cite{sliwerski2005changes}, to identify VICs. However, the V-SZZ algorithm is unable to locate VICs of VFCs that contain no deleted lines~\cite{bao2022v}, leading to missing VICs for several VFCs. To alleviate this limitation, we also adopted the developer-informed labeling method proposed by Lyu et al.~\cite{lyu2024evaluating}. In the Linux kernel repository, it is recommended that VFCs explicitly state their VICs~\cite{LinuxKernelPatches}. As shown in Figure~\ref{fig:linuxfix}, this patch states its vulnerable commit using the syntax: \texttt{Fixes: <commit-id-string>}. Leveraging this convention, we applied regular expressions to extract VICs directly from commit messages in the Linux dataset, using only these explicitly referenced commits to improve the precision of the labeled data. In contrast, due to the lack of standardized labeling practices in the FFmpeg repository, we continued to rely on the V-SZZ approach for identifying VICs in that dataset.

Using a similar syntax, we could expand our VICs collection by leveraging the developers' annotations.

\vspace{2mm}

\textbf{Data Splitting.} In total, we curated 88,178 and 993,704 commits for FFmpeg and Linux, respectively, resulting in a dataset comprising 1,081,882 commits. To the best of our knowledge, this is the largest dataset currently available for JIT-VP.
We then partitioned this dataset into training, validation, and testing sets using a time-wise approach~\cite{zeng2021deep, lyu2023chronos}, maintaining the chronological order of the data to ensure a practical evaluation that reflects continuous software development and integration~\cite{arani2024systematic} and avoid data leakage~\cite{le2019automated}. 
In particular, the commits in the training, validation, and testing sets were ordered by their commit date to reflect the evolution of open-source code~\cite{mcintosh2018fix}. We also preserved the ratio of VICs between different sets to ensure that the learning models have enough positive instances to differentiate vulnerable code from safe ones. We first selected two anchored VICs (VIC-1 and VIC-2) to divide the VIC set into 75\% for training, 5\% for validation, and 20\% for testing. Next, we incorporated the VFCs and VNCs. Specifically, VFCs and VNCs dated before VIC-1 were added to the training set, those dated after VIC-2 were assigned to the testing set, and the remaining commits were included in the validation set. These partitions of VICs, VFCs, and VNCs were utilized in all of our experiments. The detailed statistics of our datasets are presented in Table \ref{data:stat}. 

\subsection{Evaluation Metrics}
In this subsection, we present our evaluation metrics and the rationale behind our selection.

In JIT-VP, evaluation metrics commonly include ROC-AUC and F1-score~\cite{nguyen2024code, lomio2022just}. ROC-AUC measures the ability of a model to distinguish between positive and negative instances by calculating the area under the ROC curve, while the F1 score balances precision and recall. However, both metrics focus primarily on positive instances and are most effective for balanced datasets. As shown in Table \ref{data:stat}, the dataset is highly imbalanced, with significantly fewer VICs than other instances, which makes these metrics inadequate for practical evaluation. Thus, we include them only for completeness and comparison with previous studies.

To adequately assess JIT-VP in the context of data imbalance, we utilize two new evaluation metrics: Precision-Recall Area Under the Curve (PR-AUC) and Matthews Correlation Coefficient (MCC).

\textbf{PR-AUC} measures classifier performance by quantifying the area under the precision-recall curve. In JIT-VP, the main objective is to maximize the detection of vulnerable commits (recall) while minimizing false alarms (precision). Since these metrics vary with thresholds, PR-AUC provides a more reliable evaluation by integrating them across all thresholds.

\textbf{MCC} is a special case of the $\phi$ coefficient~\cite{fruchter1989j}, which measures the effectiveness of classification models over True Positives (TP), True Negatives (TN), False Positives (FP), and False Negatives (FN) as defined by the following formula: 

{\scriptsize
\begin{equation*}  
    MCC = \frac{1}{2} \cdot \left( \frac{\left(TP \cdot TN \right) - \left(FP \cdot FN \right) }{\sqrt{\left(TP + FP \right) \left(TP + FN \right) \left(TN + FP \right) \left(TN + FN \right)}} + 1\right)    
\end{equation*} 
}

MCC has been widely proven to be reliable in measuring classification results on imbalanced datasets~\cite{chicco2020advantages, zhu2020performance,le2022use}.

\begin{table}[t]
\centering
\caption{Summary of nine expert features used in the VCCFinder model.}
\label{model:vccef}
\resizebox{0.95\columnwidth}{!} {
\begin{tabular}{@{}l|l@{}}
\hline
\multicolumn{1}{c|}{\textbf{Feature Name}} & \multicolumn{1}{c}{\textbf{Description}} \\ \hline
addition                & Number of added lines                    \\
deletion               & Number of deleted lines \\
hunk\_count            & Number of hunks (code blocks) modified                  \\
kw\_x                  & Number of structural C/C++ keywords       \\
author\_contributions  & Percentage of contributions made by an author          \\
past\_changes         & Number of changes to the files in the past       \\
future\_changes       & Number of changes to the files in the future     \\
past\_different\_authors   & Number of authors modifying the files in the past\\
future\_different\_authors & Number of authors modifying the files in the future\\ \hline
\end{tabular}
}
\end{table}

\subsection{Studied Just-In-Time Vulnerability Prediction Models}
\label{sec:models}
In this subsection, we briefly describe the details of the state-of-the-art JIT-VP approaches, as well as influential JIT Defect Prediction (JIT-DP) models that are adaptable to vulnerability identification. JIT-DP models are approaches that provide early feedback to identify general defects. Since vulnerabilities can refer to ``security bugs'', JIT-DP models can be adapted to JIT-VP tasks. To ensure the correct implementation of existing studies, we selected only the models that publicly shared their replication package or had been revisited by other studies.
The summary of techniques and their utilized features is presented in Table~\ref{model:model}.

\begin{table}[t]
\centering
\caption{Summary of 14 expert features used in Just-in-time Defect Prediction models.}
\label{mode:14f}
\resizebox{0.95\columnwidth}{!} {
\begin{tabular}{@{}c|l@{}}
\hline
\textbf{Feature} & \multicolumn{1}{c}{\textbf{Description}} \\ \hline
NS      & Number of modified subsystems                              \\
ND      & Number of modified directories                             \\
NF      & Number of modified files                                   \\
Entropy & Distribution of modified code across files                 \\
LA      & Lines of code added                                        \\
LD      & Lines of code deleted                                      \\
LT      & Lines of code in a file before the change                  \\
FIX     & Whether or not the change is a defect fix (binary)         \\
NDEV    & Number of developers who modified the files                \\
AGE     & Average time interval between the last and current change  \\
NUC     & Number of unique changes to the modified files             \\ 
EXP     & Developer's experience                                     \\
REXP    & Recent developer's experience                             \\
SEXP    & Developer's experience within a subsystem                 \\ \hline
\end{tabular}
}
\end{table}

\begin{table}[t]
\centering
\caption{Summary of the studied approaches. EF is expert features. CM is commit messages. CC is commit changes.}
\label{model:model}
\begin{tabular}{@{}l|ccc|l@{}}
\hline
\multirow{2}{*}{\textbf{Models}} & \multicolumn{3}{c|}{\textbf{Features}}                          & \multirow{2}{*}{\textbf{Technique}} \\\cline{2-4}
                   & \multicolumn{1}{c|}{EF}         & \multicolumn{1}{c|}{CM}         & CC         &                  \\ \hline
\textbf{VCCFinder}~\cite{perl2015vccfinder} & \multicolumn{1}{c|}{\checkmark} & \multicolumn{1}{c|}{\checkmark}           &            & Machine Learning \\
\textbf{CodeJIT}~\cite{nguyen2024code}   & \multicolumn{1}{c|}{}           & \multicolumn{1}{c|}{}           & \checkmark & Graph-based Learning \\
\textbf{LR}~\cite{kamei2012large}        & \multicolumn{1}{c|}{\checkmark} & \multicolumn{1}{c|}{}           &            & Machine Learning \\
\textbf{TLEL}~\cite{yang2017tlel}      & \multicolumn{1}{c|}{\checkmark} & \multicolumn{1}{c|}{}           &            & Machine Learning \\                  
\textbf{DeepJIT}~\cite{hoang2019deepjit}   & \multicolumn{1}{c|}{}           & \multicolumn{1}{c|}{\checkmark} & \checkmark & Deep Learning    \\
\textbf{LAPredict}~\cite{zeng2021deep} & \multicolumn{1}{c|}{\checkmark} & \multicolumn{1}{c|}{}           &            & Machine Learning \\
\textbf{SimCom}~\cite{zhou2022simple}   & \multicolumn{1}{c|}{\checkmark} & \multicolumn{1}{c|}{\checkmark} & \checkmark & Ensemble Learning  \\
\textbf{JITFine}~\cite{ni2022best}   & \multicolumn{1}{c|}{\checkmark}           & \multicolumn{1}{c|}{\checkmark} & \checkmark & Deep Learning    \\ \hline
\end{tabular}
\end{table}

\textbf{VCCFinder}~\cite{perl2015vccfinder} employs a Support Vector Machine to distinguish VICs from other commits based on various features defined by experts (see Table~\ref{model:vccef}) and commit messages. These features fall into two categories: (i) code metrics, e.g., addition, 
which reveal notable characteristics of commit changes, and (ii) metadata, e.g., author\_contribution, which shows the author's experiences and highlights files that are frequently modified. 
Unfortunately, the VCCFinder replication package is not available. Therefore, in this study, we utilize a replicated version of VCCFinder, developed by Riom et al.~\cite{riom2021revisiting}, which is widely used in literature~\cite{nguyen2024code, chen2024improvingdatacurationsoftware}.

\textbf{CodeJIT}~\cite{nguyen2024code} is a JIT-VP approach that emphasizes code changes as key vulnerability indicators. It introduces the Code Transformation Graph (CTG), built on the Code Property Graph (CPG)~\cite{yamaguchi2014modeling}, to effectively capture information from code changes. Then, it utilizes graph neural network (GNN) models, i.e.,  RGCN, FastRGCN, and RGAT, to identify vulnerable commits. However, during the experiments, we found that CodeJIT is limited by extensive computational resources to build the CTG using the Joern tool~\cite{joern}, posing significant challenges for large-scale datasets.


\textbf{LR}~\cite{kamei2012large} is a Logistic Regression approach for JIT-DP using 14 expert features listed in Table~\ref{mode:14f}. These features represent various aspects of a commit, such as diffusion (how far this change affects the repository), size (how large this change is), purpose (whether this change fixes a bug), history (how often changes are added to the repository), and experience (how often the author changes the repository). This work popularizes the use of expert features in both JIT-VP and JIT-DP.

\textbf{TLEL}~\cite{yang2017tlel} is a two-layer ensemble learning approach. Each layer contains multiple Random Forest models combined by bagging Decision Trees. TLEL's results are highly favorable compared to ensemble learning in JIT-DP. Note that while TLEL was originally proposed for JIT-DP, it also closely resembles the most effective JIT-VP model reported by Lomio et al.~\cite{lomio2022just}.

\textbf{DeepJIT}~\cite{hoang2019deepjit} is an end-to-end deep learning model for JIT-DP. This model utilizes two Convolutional Neural Networks (CNNs) to automatically extract notable features from given commit changes and messages, respectively. The two extracted vector representations are concatenated and fed to a fully connected layer to obtain the output. DeepJIT effectively sets the stage for other deep learning-based approaches to JIT defect and vulnerability prediction~\cite{le2020deep}.

\textbf{LAPredict}~\cite{zeng2021deep} is a simple logistic regression model based on a single feature, the number of lines added, motivated by findings that this feature is the most influential among the 14 expert features proposed in Kamei et al.~\cite{kamei2012large}.

\begin{table*}[t]
\caption{Performance of the studied models in the idealized setting. Bold numbers denote the highest values.}
\centering
\renewcommand{\arraystretch}{1} 
\setlength{\tabcolsep}{6pt} 
\begin{tabular}{c|c|cccccccc|c}
\hline
 & \textbf{Metric} & \textbf{VCCFinder} & \textbf{LAPredict} & \textbf{LR} & \textbf{TLEL} & \textbf{SimCom} & \textbf{DeepJIT} & \textbf{JITFine} & \textbf{CodeJIT} & \textbf{Average} \\ 
\hline
\multirow{4}{*}{\rotatebox{90}{\textbf{FFmpeg}}} 
& \text{PR-AUC}    & 0.895  & 0.558  & 0.780  & 0.850  & 0.921  & 0.906  & \textbf{0.959}  & 0.798  & 0.833  \\
& \text{MCC}       & 0.746  & 0.337  & 0.574  & 0.701  & 0.770  & 0.638  & \textbf{0.864}  & 0.579  & 0.651  \\
& \text{F1-score}  & 0.832  & 0.373  & 0.671  & 0.800  & 0.847  & 0.759  & \textbf{0.909}  & 0.716  & 0.738  \\
& \text{ROC-AUC}   & 0.948  & 0.620  & 0.865  & 0.918  & 0.954  & 0.946  & \textbf{0.980}  & 0.838  & 0.884  \\ 
\hline
\multirow{4}{*}{\rotatebox{90}{\textbf{Linux}}}  
& \text{PR-AUC}    & 0.809  & 0.610  & 0.788  & 0.829  & \textbf{0.892}  & 0.823  & 0.885  & ---   & 0.805  \\
& \text{MCC}       & 0.447  & 0.358  & 0.558  & 0.627  & 0.658  & 0.613  & \textbf{0.716}  & ---   & 0.568  \\
& \text{F1-score}  & 0.664  & 0.458  & 0.695  & 0.752  & 0.786  & 0.735  & \textbf{0.818}  & ---   & 0.701  \\
& \text{ROC-AUC}   & 0.867  & 0.699  & 0.825  & 0.881  & 0.913  & 0.873  & \textbf{0.915}  & ---   & 0.853  \\ 
\hline
\end{tabular}
\label{eval:lab}
\end{table*}

\textbf{SimCom}~\cite{zhou2022simple} takes advantage of ensemble learning for JIT-DP. This is a combination of two models, Sim and Com. Sim is a simple decision tree trained on 14 change features in Table~\ref{mode:14f}, while Com is a CNN-based model similar to DeepJIT~\cite{hoang2019deepjit}. Each model predicts the given commit independently. Then, the output is the average of the two predictions. The encouraging results of SimCom show the potential of a mixture of models to JIT-DP as well as JIT-VP.

\textbf{JITFine}~\cite{ni2022best} integrates both traditional expert features and deep features extracted from CodeBERT~\cite{feng2020codebert} for JIT-DP. Specifically, this work uses CodeBERT to embed commit changes and messages. Then, the embedding vector, along with expert features, would be fed to an integrated feature learning network. The output of this procedure is the commit representation that will be utilized to predict defective commits.

\subsection{Implementation Details}
\label{implemetation}
We reused the original implementations of the studied models, training and evaluating them on our dataset. For VCCFinder, we used the implementation by Riom et al.~\cite{riom2021revisiting} due to the unavailability of the original. All models were run with parameters from their replication packages on a NVIDIA A100 GPU (40GB VRAM), 250GB RAM, and a 32-core Intel Xeon CPU (2.90GHz) running Red Hat Enterprise Linux 9.3 and Java 1.8.0\_241.

\section{Experiments and Results}
\label{sec:eval}

\subsection{RQ$_{1}$: JIT-VP Performances in the Idealized Setting}
\label{sec:rq1}

\subsubsection{Experiment Design}
To answer RQ$_{1}$, we conducted experiments by training and evaluating the state-of-the-art JIT-VP techniques, as described in Section~\ref{sec:models}, within the \textbf{idealized setting} outlined in Section~\ref{sec:data_col_limits}. The evaluation of CodeJIT~\cite{nguyen2024code} was limited to the FFmpeg dataset due to the prohibitive computational cost of constructing graphs for all commits in the Linux kernel dataset, as detailed in Section~\ref{sec:models}.

\subsubsection{Results} Table~\ref{eval:lab} presents the results of our RQ$_{1}$ experiments. Overall, the models exhibit strong performance in the idealized setting across the Linux and FFmpeg datasets.
Specifically, the models achieve average PR-AUC values of 0.833 and 0.805, and the highest PR-AUC values of 0.959 and 0.892 for FFmpeg and Linux, respectively. Under this setting, the performance values suggest that the current JIT-VP models can be useful for developers~\cite{neuhaus2007predicting,shin2013can}.

A closer comparison with prior work reveals some variations. Notably, CodeJIT~\cite{nguyen2024code}, the latest JIT-VP study, reported an F1-score of 0.74 in the development process setting, where commits are ordered chronologically. In our replication, the model achieves a slightly lower F1-score of 0.716.
Other models, such as JITFine and DeepJIT, show higher performance on our dataset than the numbers reported in the CodeJIT study. For instance, JITFine achieves F1-scores of 0.909 and 0.818 on our FFmpeg and Linux datasets, respectively, but it could only achieve an F1-score of 0.67 in the CodeJIT's experiments. 

These variations can be caused by the differences in the experimental settings. Although our study adopts a within-project setting, where the training and testing datasets come from the same project, CodeJIT employs a cross-project setting due to its limited dataset size. Our Linux dataset has 4,614 VICs and 4,042 VFCs, providing a more extensive training set that allows for evaluation within the project. In contrast, CodeJIT's dataset consists of only 783 VICs and 780 VFCs, necessitating a cross-project approach to ensure sufficient data.

Despite these differences, our findings remain consistent with the key previous findings in~\cite{nguyen2024code} that JIT-VP techniques exhibit impressive performance in the idealized setting when trained and evaluated only on VICs and VFCs. Accordingly, these results can provide a reliable baseline for comparison with the realistic setting in RQ$_{2}$.

\begin{tcolorbox}
\textbf{Answer to RQ$_{1}$:} JIT-VP models achieve strong performances with PR-AUC of 0.8+ on average and 0.959 at best in the idealized setting across the Linux and FFmpeg datasets. Our results align with prior findings.
\end{tcolorbox}

\begin{table*}[t]
\caption{Performance of the studied models trained in the idealized setting and tested in the realistic setting. Bold numbers indicate the highest performance per metric. The \textbf{Ideal} column shows average RQ$_{1}$ results (Table~\ref{eval:lab}), while the \textbf{Drop} column represents the percentage decline from the idealized setting.}
\centering
\renewcommand{\arraystretch}{1} 
\setlength{\tabcolsep}{6pt} 
\begin{tabular}{@{}c|c|cccccccc|c|c|c@{}}
\hline
& \textbf{Metric}
   &
  \textbf{VCCFinder} &
  \textbf{LAPredict} &
  \textbf{LR} &
  \textbf{TLEL} &
  \textbf{SimCom} &
  \textbf{DeepJIT} &
  \textbf{JITFine} &
  \textbf{CodeJIT} &
  \textbf{Average} &
  \textbf{Ideal} &
  \textbf{Drop} \\ \hline
\multirow{4}{*}{\rotatebox{90}{\textbf{FFmpeg}}} & \text{PR-AUC}   & 0.055 & 0.045 & 0.076          & \textbf{0.099} & 0.081 & 0.094 & 0.079          & 0.074 & 0.076 & 0.833 & 90.93\% \\
                                 & \text{MCC}       & 0.032 & 0.059 & 0.132          & 0.089          & 0.063 & 0.036 & \textbf{0.135} & 0.125 & 0.078 & 0.651 & 88.01\% \\
                                 & \text{F1-score} & 0.057 & 0.085 & 0.121          & 0.077          & 0.066 & 0.056 & \textbf{0.149} & 0.122 & 0.087 & 0.738 & 88.18\% \\
                                 & \text{ROC-AUC}  & 0.631 & 0.590 & 0.695          & \textbf{0.719} & 0.698 & 0.691 & 0.711          & 0.621 & 0.676 & 0.884 & 23.45\% \\ \hline
\multirow{4}{*}{\rotatebox{90}{\textbf{Linux}}}  & \text{PR-AUC}   & 0.018 & 0.011 & \textbf{0.024} & 0.019          & 0.014 & 0.014 & 0.019          & ---     & 0.017 & 0.805 & 97.85\% \\
                                 & \text{MCC}       & 0.005 & 0.030 & 0.061          & 0.061          & 0.031 & 0.019 & \textbf{0.135} & ---     & 0.049 & 0.568 & 91.42\% \\
                                 & \text{F1-score} & 0.011 & 0.021 & 0.027          & 0.025          & 0.016 & 0.014 & \textbf{0.149} & ---     & 0.038 & 0.701 & 94.63\% \\
                                 & \text{ROC-AUC}  & 0.653 & 0.591 & 0.733          & \textbf{0.750} & 0.669 & 0.661 & 0.711          & ---     & 0.681 & 0.853 & 20.19\% \\ \bottomrule
\end{tabular}
\label{eval:test}
\end{table*}

\begin{table*}[t]
\caption{Performance of the studied models both trained and tested in the realistic setting. The bold numbers indicate the highest performance for each metric. The \textbf{Ideal} column shows average RQ$_{1}$ results (Table~\ref{eval:lab}), while the \textbf{Drop} column represents the percentage decline from the idealized setting.}
\centering
\renewcommand{\arraystretch}{1} 
\setlength{\tabcolsep}{6pt} 
\begin{tabular}{@{}c|c|cccccccc|c|c|c@{}}
\hline
 &
   \textbf{Metric} &
  \textbf{VCCFinder} &
  \textbf{LAPredict} &
  \textbf{LR} &
  \textbf{TLEL} &
  \textbf{SimCom} &
  \textbf{DeepJIT} &
  \textbf{JITFine} &
  \textbf{CodeJIT} &
  \textbf{Average} &
  \textbf{Ideal} &
  \textbf{Drop} \\ \hline
\multirow{4}{*}{\rotatebox{90}{\textbf{FFmpeg}}} & \text{PR-AUC}    & 0.071 & 0.041 & 0.093          & 0.112          & \textbf{0.134} & 0.082 & 0.111 & 0.079 & 0.091 & 0.833 & 89.14\% \\
                                 & \text{MCC}       & 0.122 & 0.067 & 0.192          & 0.169          & \textbf{0.226} & 0.138 & 0.161 & 0.135 & 0.151 & 0.651 & 76.78\% \\
                                 & \text{F1-score}  & 0.132 & 0.086 & 0.176          & 0.130          & \textbf{0.231} & 0.150 & 0.156 & 0.142 & 0.150 & 0.738 & 79.63\% \\
                                 & \text{ROC-AUC}   & 0.688 & 0.591 & 0.769          & 0.795          & \textbf{0.809} & 0.746 & 0.790 & 0.721 & 0.738 & 0.884 & 16.43\% \\ \hline
\multirow{4}{*}{\rotatebox{90}{\textbf{Linux}}}  & \text{PR-AUC}   & 0.013 & 0.010 & 0.023          & 0.027          & \textbf{0.031} & 0.005 & 0.005 & ---     & 0.016 & 0.805 & 97.99\% \\
                                 & \text{MCC}       & 0.030 & 0.028 & 0.070          & \textbf{0.081} & 0.073          & 0.000 & 0.000 & ---     & 0.040 & 0.568 & 92.92\% \\
                                 & \text{F1-score} & 0.036 & 0.025 & \textbf{0.039} & 0.038          & 0.034          & 0.011 & 0.011 & ---     & 0.028 & 0.701 & 96.04\% \\
                                 & \text{ROC-AUC}  & 0.588 & 0.591 & 0.746          & \textbf{0.787} & 0.779          & 0.497 & 0.497 & ---     & 0.641 & 0.853 & 24.90\% \\ \bottomrule
\end{tabular}
\label{eval:wild}
\end{table*}

\subsection{RQ$_{2}$: JIT-VP Performances in the Realistic Setting}
\label{sec:rq2}

\subsubsection{Experimental Design}
To answer the RQ$_{2}$, we conducted experiments using the \textbf{realistic setting} described in Section~\ref{sec:data_col_limits}, where the evaluation dataset included all types of commits, namely VICs, VFCs, and VNCs.
Specifically, we performed two experiments: (\textit{i}) trained the models on idealized data (as described in Section~\ref{sec:rq1}) and tested them on the realistic data and (\textit{ii}) trained and tested the models directly on the realistic data.
Similar to RQ$_{1}$, CodeJIT experiments on the Linux kernel dataset were excluded due to the prohibitive computational cost associated with code graph construction.

\begin{figure}
    \centering
    \includegraphics[width=\linewidth]{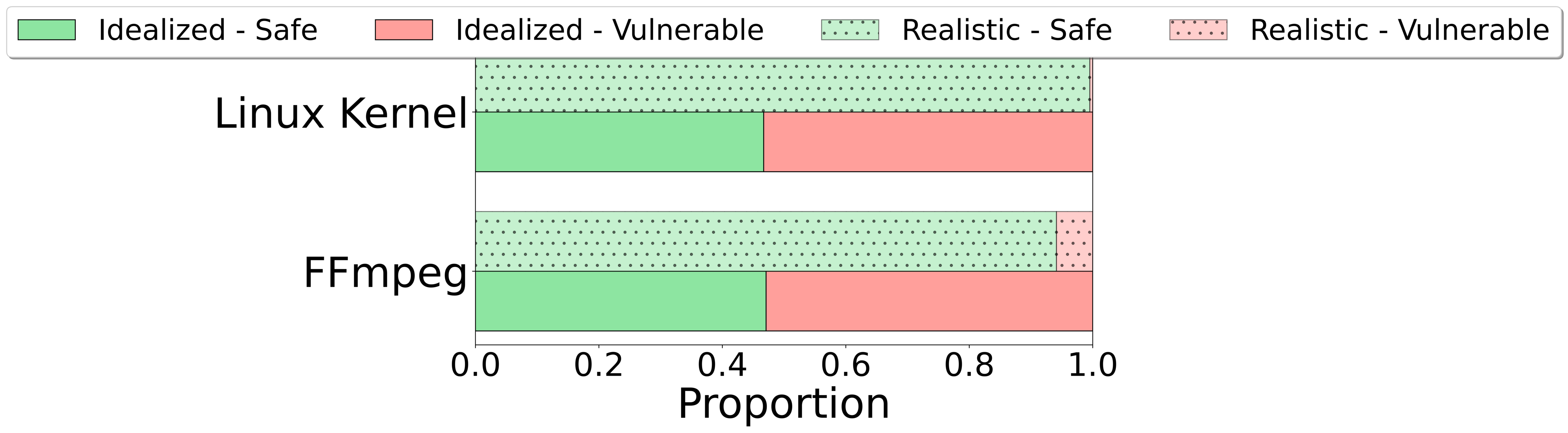}
    \caption{Comparison of the ratio of safe to vulnerable commits in the testing data between the idealized and realistic settings.}
    \label{fig:ratio}
    \vspace{0mm}
\end{figure}

\subsubsection{Results}
The models trained in the idealized setting experience a significant performance drop when tested with realistic data (see Table~\ref{eval:test}). The PR-AUC metric drops drastically across all models, i.e., by 90.93\% on average (from 0.833 to 0.076) on FFmpeg and by 97.85\% (from 0.805 to 0.017) on Linux.
We also observe similar performance declines in terms of MCC and F1-score.
In contrast, the average declines in the ROC-AUC are less significant at 23.45\% and 20.19\% on FFmpeg and Linux, respectively. However, it is important to note that the realistic setting introduces a significantly imbalanced distribution between the two classes (see Table~\ref{data:stat}), making ROC-AUC a less suitable evaluation metric than MCC and PR-AUC.
These findings imply that models trained in the idealized setting fail to generalize effectively when exposed to realistic scenarios with a more imbalanced distribution between vulnerable and safe commits. This result is reasonable, as the training data for these models do not include VNCs, making it challenging for JIT models to distinguish these commits from VICs. To address this issue, we attempted to train JIT models directly in a realistic setting, where the training data include VICs, VFCs, and VNCs.

Unfortunately, even in the realistic setting, the models still consistently demonstrate low performance (see Table~\ref{eval:wild}). For instance, PR-AUC remains low across all models, achieving the highest values of 0.134 on FFmpeg and 0.031 on the Linux kernel when using the SimCom model.
Notably, DeepJIT and JITFine, the two best-performing techniques in the idealized setting, even exhibit an MCC of 0 and PR-AUC of 0.005 on Linux.
This poor performance can be attributed to the highly imbalanced nature of the dataset, which causes these deep learning-based models to be biased toward the majority class. Consequently, they tend to predict all instances as belonging to this class with high probability. This finding highlights the limitations of deep learning-based approaches in the realistic setting.

We find that drastic performance declines are primarily caused by a severe data imbalance, as illustrated in Figure~\ref{fig:ratio}. In the idealized setting, the ratio between vulnerable and safe commits is relatively balanced, with a ratio of 1:0.89 in FFmpeg and 1:0.88 in Linux. However, in the realistic setting, this imbalance is significantly more pronounced. For instance, the ratio of vulnerable to safe commits in FFmpeg is only 1:17, while in Linux, it is even more extreme at 1:127. Consequently, if JIT models are trained in the realistic setting, the overwhelming presence of negative instances shifts the model learning parameters toward predicting 0. However, the primary objective of JIT-VP is to identify positive cases; i.e., the models receive more rewards for predicting vulnerabilities. This confusion caused by imbalanced data poses a fundamental challenge for learning-based approaches in effectively differentiating between the two classes.

\begin{table*}[t]
\centering
\renewcommand{\arraystretch}{1} 
\setlength{\tabcolsep}{6pt} 
\caption{Performance of the random sampling methods. The bold numbers indicate the highest performance for each metric. The \textbf{w/o Sampling} column shows the original average results of JIT-VP in the realistic setting in RQ$_{2}$ without random sampling techniques based on Table~\ref{eval:wild}. }
\label{eval:sample}
\begin{tabular}{@{}c|c|ccccccc|c|c@{}}
\hline
 & \textbf{Metric}
   & 
  \textbf{VCCFinder} &
  \textbf{LAPredict} &
  \textbf{LR} &
  \textbf{TLEL} &
  \textbf{SimCom} &
  \textbf{DeepJIT} &
  \textbf{JITFine} &
  \textbf{Average} &
  \textbf{w/o Sampling} \\ \hline
\multirow{4}{*}{\textbf{FFmpeg-RUS}} & \text{PR-AUC}   & 0.084 & 0.045 & 0.088          & 0.118          & 0.104 & \textbf{0.237} & 0.124          & 0.114 & 0.092 \\
                                     & \text{MCC}      & 0.115 & 0.060 & \textbf{0.171} & 0.156          & 0.004 & 0.004          & 0.142          & 0.093 & 0.153 \\
                                     & \text{F1-score} & 0.099 & 0.087 & \textbf{0.167} & 0.130          & 0.051 & 0.051          & 0.112          & 0.100 & 0.152 \\
                                     & \text{ROC-AUC}  & 0.722 & 0.590 & 0.741          & \textbf{0.774} & 0.767 & 0.654          & 0.767          & 0.716 & 0.741 \\ \hline
\multirow{4}{*}{\textbf{FFmpeg-ROS}} & \text{PR-AUC}   & 0.076 & 0.045 & 0.092          & 0.116          & 0.108 & 0.026          & \textbf{0.127} & 0.084 & 0.092 \\
                                     & \text{MCC}      & 0.118 & 0.060 & 0.179          & \textbf{0.181} & 0.031 & 0.031          & 0.157          & 0.108 & 0.153 \\
                                     & \text{F1-score} & 0.133 & 0.087 & \textbf{0.176} & 0.165          & 0.053 & 0.051          & 0.122          & 0.112 & 0.152 \\
                                     & \text{ROC-AUC}  & 0.664 & 0.590 & 0.738          & 0.771          & 0.756 & 0.500          & \textbf{0.790} & 0.687 & 0.741 \\ \hline
\multirow{4}{*}{\textbf{Linux-RUS}}  & \text{PR-AUC}   & 0.067 & 0.010 & 0.023          & 0.026          & 0.026 & \textbf{0.160} & 0.122          & 0.062 & 0.016 \\
                                     & \text{MCC}      & 0.103 & 0.027 & 0.068          & 0.075          & 0.002 & 0.002          & \textbf{0.151} & 0.061 & 0.040 \\
                                     & \text{F1-score} & 0.084 & 0.024 & 0.035          & 0.033          & 0.012 & 0.012          & \textbf{0.109} & 0.044 & 0.028 \\
                                     & \text{ROC-AUC}  & 0.722 & 0.591 & 0.728          & \textbf{0.785} & 0.775 & 0.769          & 0.780          & 0.736 & 0.641 \\ \bottomrule
\end{tabular}
\end{table*}

\begin{tcolorbox}
\textbf{Answer to RQ$_{2}$:} The effectiveness of the state-of-the-art JIT-VP significantly declines in the realistic setting, exhibiting reductions of 89\% -- 98\% in PR-AUC across the FFmpeg and Linux datasets.
\end{tcolorbox}

\subsection{RQ$_{3}$: Exploration of Data Imbalance Mitigation for JIT-VP}
\label{sec:rq3}

\subsubsection{Experiment Design}
In RQ$_{3}$, we conducted experiments to explore the mitigation of the data imbalance existing in the realistic setting for JIT-VP, as identified in RQ$_{2}$. We also applied all these techniques during the training phase and \textit{not} in the validation/testing phases. The studied techniques to address the data imbalance are presented below.

\textbf{Random Under-Sampling (RUS)}.  This technique modifies the class distribution by randomly removing samples from the majority class (i.e., safe commits). The data are sampled to reach the ratio of 1:1 positive-to-negative.

\textbf{Random Over-Sampling (ROS)}. 
    This technique modifies the class distribution by augmenting the minority class through random duplication. The data are sampled to reach the ratio of 1:1 positive-to-negative. Due to the large number of commits in Linux, we applied ROS exclusively to the FFmpeg dataset.

\textbf{Synthetic Minority Oversampling Technique (SMOTE)~\cite{fernandez2018smote}}. This technique generates synthetic samples for the minority class, thus enhancing model learning without simple duplication. The positive-to-negative ratio of data is 1:1. Since SMOTE is applicable only to tabular data, we applied it exclusively to three techniques: LAPredict, LR, and TLEL.

\textbf{One-Sided Selection (OSS)~\cite{kubat1997addressing}}. This technique removes redundant majority-class instances, reduces noise, and improves the quality of the training data. Similar to SMOTE, this technique is applicable only to tabular data, so we also applied it exclusively to three techniques: LAPredict, LR, and TLEL, and keep the positive-to-negative ratio of the data 1:1.

\textbf{Focal Loss (FL)}. Focal Loss~\cite{lin2017focal}, a widely used method to address data imbalance, including in software engineering tasks~\cite{zhou2023devil}.
Unlike sampling-based techniques, FL modifies the loss function to reduce the influence of well-classified instances and emphasize hard-to-classify samples. By dynamically scaling the loss contribution based on prediction confidence, FL enables models to focus on misclassified examples. Formally, FL is defined as: $FL(p_t) = -\alpha_t (1 - p_t)^\gamma \log(p_t)$, where \( p_t \) is the predicted probability of the class of interest. \( \alpha_t \) is a factor for class imbalance (\( \alpha_t \in [0,1] \)) that assigns importance to different classes. \( \gamma \) is a focusing parameter (\( \gamma \geq 0 \)), which controls the down-weighting of well-classified samples. We trained DeepJIT and SimCom using FL with \(\gamma = 2\) and compared their performance with models trained using standard Cross-Entropy Loss.

\subsubsection{Results} Tables~\ref{eval:sample}, \ref{eval:advsample}, and \ref{eval:focalloss} present the results for the studied random sampling techniques (RUS, ROS), advanced sampling techniques (SMOTE, OSS) and FL, respectively.

\textbf{Random Sampling.} From Table~\ref{eval:sample}, we observe that RUS can marginally improve JIT-VP models in terms of PR-AUC. Specifically, RUS slightly increases the average PR-AUC of these models from 0.092 to 0.114 on the FFmpeg dataset and from 0.016 to 0.062 on the Linux dataset. This suggests that while RUS offers improvements on both datasets, the overall effectiveness of JIT-VP techniques remains considerably low.

In contrast, ROS appears to have a slightly adverse effect on PR-AUC. The average PR-AUC of the models decreases marginally from 0.092 to 0.084 on FFmpeg, indicating that oversampling the minority class does not necessarily translate to improved model performance. This suggests that ROS may introduce data redundancy or increase overfitting, reducing the model's generalization capability.

Notably, RUS and ROS lead to a significant reduction in MCC for these models on FFmpeg. The average MCC drops from 0.153 to 0.093 with RUS and further decreases to 0.061 with ROS. This indicates a decline in the models' ability to effectively differentiate between vulnerable and non-vulnerable commits under these sampling strategies. Although RUS slightly improves MCC on Linux, increasing it from 0.040 to 0.061, the overall improvement remains marginal.

These results highlight the mixed impacts of RUS and ROS on JIT-VP techniques. Even in cases where improvements are observed, they are usually minor and inconsistent across datasets. These results highlight the limited effectiveness of random sampling techniques in improving JIT-VP models.

\begin{table}[t]
\caption{Average results of advanced sampling methods on expert feature data. The bold numbers indicate the highest performance for each metric. The w/o Sampling column shows the original average results of JIT-VP on the realistic setting without advanced sampling techniques based on Table~\ref{eval:wild}.}
\centering
\renewcommand{\arraystretch}{1} 
\setlength{\tabcolsep}{6pt}
\resizebox{\columnwidth}{!} {
\begin{tabular}{@{}c|c|ccc|c|c@{}}
\hline
                                       &  \textbf{Metric}               & \textbf{LAPredict} & \textbf{LR} & \textbf{TLEL}  & \textbf{Average} & \textbf{w/o sampling} \\ \hline
\multirow{4}{*}{\textbf{\rotatebox{90}{SMOTE}}} & \text{PR-AUC} & 0.045              & 0.094       & \textbf{0.112} & 0.084            & 0.082              \\
 & \text{MCC}      & 0.060 & \textbf{0.181} & 0.156          & 0.132 & 0.142 \\
 & \text{F1-score} & 0.087 & \textbf{0.176} & 0.132          & 0.132 & 0.131 \\
 & \text{ROC-AUC}  & 0.590 & 0.747          & \textbf{0.765} & 0.701 & 0.718 \\ \hline
\multirow{4}{*}{\textbf{\rotatebox{90}{OSS}}}   & \text{PR-AUC} & 0.045              & 0.090       & \textbf{0.103} & 0.080            & 0.082              \\
 & \text{MCC}      & 0.060 & 0.176          & \textbf{0.177} & 0.138 & 0.142 \\
 & \text{F1-score} & 0.087 & \textbf{0.177} & 0.171          & 0.145 & 0.131 \\
 & \text{ROC-AUC}  & 0.590 & 0.739          & \textbf{0.768} & 0.699 & 0.718 \\ \bottomrule
\end{tabular}
}
\label{eval:advsample}
\end{table}

\textbf{Advanced Sampling.} From Table~\ref{eval:advsample}, we observe that even advanced sampling methods such as SMOTE and OSS exhibit limited effectiveness in mitigating data imbalance for JIT-VP models. While SMOTE slightly improves the average MCC from 0.082 to 0.084 and the average PR-AUC from 0.132 to 0.142, these gains are much smaller than the overall performance drop observed in the realistic setting. Moreover, OSS tends to have an even more adverse effect on the JIT-VP performance. For instance, with OSS, the average PR-AUC drops to 0.080, and the MCC decreases to 0.138.
The minor improvements observed with SMOTE and the performance decline with OSS suggest that these techniques alone are not sufficient to meaningfully improve JIT-VP.

\begin{table}[t]
\caption{Average results of the Focal Loss. The bold numbers indicate the highest performance for each metric. The w/ CELoss and w/ FLoss columns reveal the results of models that adopt CELoss based on Table~\ref{eval:wild} and FLoss, respectively.}
\renewcommand{\arraystretch}{1} 
\setlength{\tabcolsep}{6pt}
\centering
\begin{tabular}{@{}c|cc|cc@{}}
\toprule
\multirow{2}{*}{\textbf{Metric}} & \multicolumn{2}{c|}{\textbf{w/ FLoss}} & \multicolumn{2}{c}{\textbf{w/ CELoss}}                  \\ \cmidrule(l){2-5} 
                                  & \textbf{SimCom}   & \textbf{DeepJIT}   & \textbf{SimCom} & \textbf{DeepJIT} \\ \midrule
PR-AUC   & \textbf{0.110} & 0.018 & {\textbf{0.134}} & 0.082 \\
MCC      & 0.000          & 0.000 & {\textbf{0.226}} & 0.138 \\
F1-score & 0.000          & 0.000 & {\textbf{0.231}} & 0.150 \\
ROC-AUC  & \textbf{0.769} & 0.334 & {\textbf{0.809}} & 0.746 \\ \bottomrule
\end{tabular}
\label{eval:focalloss}
\vspace{0mm}
\end{table}

\textbf{Focal Loss} (FL), as shown in Table~\ref{eval:focalloss}, also provides no tangible benefits and, in some cases, even worsens performance. Notably, DeepJIT and SimCom experience complete performance collapse, with MCC and F1-scores dropping to 0.000, while PR-AUC also decreases (0.110 for SimCom and 0.108 for DeepJIT). These drastic declines suggest that FL, although effective in traditional classification tasks, struggles in the context of JIT-VP.

One possible explanation for this failure is the extreme imbalance between vulnerable and non-vulnerable commits in the JIT-VP datasets. While focal loss is designed to address class imbalance by assigning a greater weight to hard-to-classify samples, it may not be well suited for JIT-VP due to the overwhelming dominance of non-vulnerable commits. In this scenario, FL could lead to ineffective gradient updates, where the models fail to learn meaningful patterns from the minority class, ultimately resulting in model collapse.

\begin{tcolorbox}
\textbf{Answer to RQ$_{3}$:} Data imbalance mitigation techniques, including data sampling and focal loss, yield only marginal performance improvements for JIT-VP approaches across the FFmpeg and Linux datasets, highlighting the persistent challenges posed by class imbalance in the realistic setting of this task.
\end{tcolorbox}

\section{Discussions}
\label{sec:discussion}
\subsection{Implications}
Based on the insights and findings in Section~\ref{sec:eval}, this section discusses the implications of our study and provides suggestions for future research directions.

\textbf{Realistic evaluation settings are crucial for JIT-VP}.
We have demonstrated significant performance gaps of up to 98\% between the idealized and realistic settings.
In the realistic setting, not only does the performance of all the models drop, but the model performance rankings are also impacted. For instance, JITFine was the best model in the idealized setting, while TLEL took the top spot in the realistic setting.
For researchers, it is important to choose the realistic setting over the idealized one moving forward; otherwise, the JIT-VP model comparison can be biased, resulting in potentially misleading conclusions.
For practitioners, it is recommended to exercise caution when considering the results reported in the idealized setting for adopting JIT-VP models in practical applications.
We propose that there should be more joint efforts between academia and industry to better align JIT-VP evaluations with practical software development environments.

\textbf{Just-in-time vulnerability prediction: Are we there yet?}
Unfortunately, our answer to this question is still no. Our findings have shed light on the true effectiveness of JIT-VP models when switching from the commonly used idealized setting in the literature to the realistic setting.
The experiments reveal that in this realistic setting, these models actually exhibit low performance (e.g., only 0.09 and 0.016 in PR-AUC) on FFmpeg and Linux, respectively. These underperformed results imply that most of current predictions are missed vulnerabilities or false alarms.
Consequently, existing models are unlikely to be ready for practical use. 
This calls for the development of more effective JIT-VP models capable of handling the complexities inherent in real-world settings.
We also highlight that one of the primary limitations to the effectiveness of JIT-VP models in the realistic setting is the significant imbalance between vulnerable and safe commits.

\textbf{Addressing the data imbalance in the realistic JIT-VP setting requires customized techniques}.
We show that many widely used techniques for mitigating data imbalance are insufficient to address this issue effectively for JIT-VP.
Most techniques yield marginal improvements in model performance under realistic settings. Although certain methods, such as applying RUS to the DeepJIT model, demonstrate notable gains (e.g., PR-AUC increases from 0.114 to 0.237 on FFmpeg and from 0.062 to 0.160 on Linux, as shown in Table~\ref{eval:sample}), these improvements are not consistently observed in other existing approaches. This lack of reproducibility suggests that such enhancements may be model- or dataset-specific, limiting their broader applicability in real-world vulnerability prediction.
Hence, it is essential to develop novel techniques specifically tailored to the unique characteristics of JIT-VP. Such advancements may require a more in-depth analysis of the data imbalance, considering its underlying causes and impact on predictive performance, which is beyond the scope of our study.
Recent advances in data augmentation/generation~\cite{nong2024vgx,nong2023vulgen,le2024mitigating,lecong2025toward} could help increase the number of vulnerable commits. However, adapting these techniques, designed for complete source code, to partial code changes requires a deep understanding of vulnerability semantics in code commits.

\subsection{Threats to Validity}

\textbf{Threats to construct validity}. To assess the practical implications of existing JIT-VP approaches, we introduce a large-scale, real-world dataset. Our dataset is curated using the state-of-the-art weakness source identification techniques (V-SZZ~\cite{bao2022v}, developer-informed labels~\cite{lyu2024evaluating}) alongside a heuristic labeling approach that marks all VFCs and VNCs as safe. One potential concern is that SZZ-based methods may introduce false positives~\cite{rosa2021evaluating}. To mitigate this, we applied strict filters (Section~\ref{sec:data_collection}) to exclude unrelated commits.
Another concern is that the heuristic labeling approach may fail to capture latent vulnerabilities~\cite{le2024latent}. While we acknowledge the elusive nature of VICs, accurately identifying them remains an open challenge. Given that safe commits significantly outnumber vulnerable ones, a small number of mislabeled instances is unlikely to substantially impact the overall empirical findings.

\textbf{Threats to internal validity}. We reused the implementations of the models, ensuring alignment with their original papers. Our experimental results align closely with the previous research findings (Section~\ref{sec:rq1}), strengthening the reliability of our implementations. To facilitate future research and reproducibility, we will release our replication package~\cite{replication}.

\textbf{Threats to external validity}. Our study focuses on widely used open-source C/C++ projects. While FFmpeg and the Linux kernel serve as representative benchmarks, our findings may not generalize to other programming languages or proprietary software. Future work should explore different ecosystems to assess broader applicability.

\section{Conclusion}
\label{sec:conclusion}

We have conducted an empirical assessment of JIT-VP performance in both controlled lab settings and real-world environments. Our primary contributions include a novel dataset for JIT-VP and a comprehensive evaluation of model effectiveness in different settings. Our findings show that JIT-VP models perform well in the idealized setting, but their efficacy significantly declines when deployed in practical scenarios. This gap highlights the challenges of translating experimental success into real-world applicability.

Furthermore, we have investigated various techniques to mitigate data imbalance, an issue that arises when adopting JIT-VP in realistic scenarios. However, our results indicate that these techniques offer only marginal improvements in practical performance.
Our findings inform the current bottleneck with JIT-VP and hopefully inspire future research to make breakthroughs in this important task of ensuring software security.
In the future, we plan to extend our investigations to enable reliable just-in-time prediction for other vulnerability management tasks~\cite{le2021deepcva,le2022survey,zhang2023survey}.


\IEEEtriggeratref{68}
\bibliographystyle{IEEEtran}
\bibliography{main}

\end{document}